\documentclass[aps,twocolumn]{revtex4}
\usepackage{amsmath,amssymb}
\begin{document}
\title{Modified Friedmann Equations from Tsallis Entropy}
\author{Ahmad Sheykhi\footnote{asheykhi@shirazu.ac.ir}}
\address{Physics Department and Biruni Observatory, College of
Sciences, Shiraz University, Shiraz 71454, Iran\\
Research Institute for Astronomy and Astrophysics of Maragha
(RIAAM), P.O. Box 55134-441, Maragha, Iran}

 \begin{abstract}
It was shown by Tsallis and Cirto that thermodynamical entropy of
a gravitational system such as black hole must be generalized to
the non-additive entropy, which is given by $S_h=\gamma
A^{\beta}$, where $A$ is the horizon area and $\beta$ is the
nonextensive parameter \cite{Tsa}. In this paper, by taking the
entropy associated with the apparent horizon of the
Friedmann-Robertson-Walker (FRW) Universe in the form of Tsallis
entropy, and assuming the first law of thermodynamics,
$dE=T_hdS_h+WdV$, holds on the apparent horizon, we are able to
derive the corresponding Friedmann equations describing the
dynamics of the universe with any spatial curvature. We also
examine the time evolution of the total entropy and show that the
generalized second law of thermodynamics is fulfilled in a region
enclosed by the apparent horizon. Then, modifying the emergence
proposal of gravity proposed by Padmanabhan and calculating the
difference between the surface degrees of freedom and the bulk
degrees of freedom in a region of space, we again arrive at the
modified Friedmann equation of the FRW Universe with any spatial
curvature which is the same as one obtained from the first law of
thermodynamics. {We also study the cosmological consequences of
Tsallis cosmology. Interestingly enough, we find that this model
can explain simultaneously the late time acceleration in the
universe filled with pressureless matter without invoking dark
energy, as well as the early deceleration. Besides, the age
problem can be circumvented automatically for an accelerated
universe and is estimated larger than $3/2$ age of the universe in
standard cosmology. Taking $\beta=2/5$, we find the age of the
universe ranges as $13.12$ Gyr $< t_0 < 16.32 $ Gyr, which is
consistent with recent observations. Finally, using the Jeans's
analysis, we comment, in brief, on the density perturbation in the
context of Tsallis cosmology and found that the growth of energy
differs compared to the standard cosmology.}

\end{abstract}
 \maketitle

 \newpage
\section{Introduction\label{Intro}}
Although gravity is the most universal forces of nature,
understanding its origin has been a mystery for a long time.
Einstein believed that gravity is just the spacetime curvature and
regarded it as an emergent phenomenon which describes the dynamics
of spacetime. In the past decades a lot of attempts have been done
to disclose the nature of gravity. A great step in this direction
put forwarded by Jacobson \cite{Jac} who studied thermodynamics of
spacetime and showed explicitly that Einstein's equation of
general relativity is just an equation of state for the spacetime.
Combining the Clausius relation $\delta Q=T\delta S$, together
with the entropy expression, he derived Einstein field equations.
This derivation is of great importance because it confirms that
the Einstein field equations is nothing but the first law of
thermodynamics for the spacetime. Following Jacobson, a lot of
studies have been carried out to disclose the deep connection
between gravity and thermodynamics \cite{Elin,Cai1,Pad}. The
studies were also generalized to the cosmological setups
\cite{Cai2,Cai3,CaiKim,Fro,verlinde,Cai4, CaiLM,Shey1,Shey2},
where it has been shown that the Friedmann equation of
Friedmann-Robertson-Walker (FRW) universe can be written in the
form of the first law of thermodynamics on the apparent horizon.
Although Jacobson's derivation is logically clear and
theoretically sound, the statistical mechanical origin of the
thermodynamic nature of general relativity remains obscure.

In $2010$ Verlinde \cite{Ver} put forwarded the next great step
toward understanding the nature of gravity who claimed that
gravity is not a fundamental force and can be interpreted as an
entropic force caused by changes of entropy associated with the
information on the holographic screen. Verlinde's proposal is
based on two principles, namely the holographic principle and the
equipartition law of energy. Using these principles he derived the
Newton's law of gravitation, the Poisson equation and in the
relativistic regime the Einstein field equations \cite{Ver} (see
also \cite{Pad0}). The investigation on the entropic origin of
gravity have been extended in different setups
\cite{Cai5,Other,newref,sheyECFE,Ling,Modesto,Yi,Sheykh2} and
references therein). Although Verlinde's proposal has changed our
understanding on the origin and nature of gravity, but it
considers the gravitational field equations as the equations of
emergent phenomenon and leave the spacetime as a pre-existed
background geometric manifold. A new perspective towards emergence
of spacetime dynamics was suggested in $2012$ by Padmanabhan
\cite{PadEm}. He argued that the spatial expansion of our Universe
can be regarded as the consequence of emergence of space and
\textit{the cosmic space is emergent as the cosmic time
progresses}. By calculating the difference between the number of
degrees of freedom in the bulk and on the boundary, Padmanabhan
\cite{PadEm} derived the Friedmann equation of the flat FRW
Universe. This proposal was also applied for deriving the
Friedmann equations of a higher dimensional FRW Universe in
Einstein, Gauss-Bonnet and more general Lovelock cosmology
\cite{CaiEm,Yang}. By modification the Padmanabhan's proposal, one
may extract the Friedmann equation of FRW Universe with any
spatial curvature not only in Einstein gravity, but also in
Gauss-Bonnet and more general Lovelock gravity \cite{Sheyem}. The
novel idea of Padmanabhan has got a lot of attentions in the
literatures \cite{FQ,Ling2,Sheyem2,Sheyem3,FF,Eu,Wen}.

It is important to note that in order to rewrite the Friedmann
equations, in any gravity theory, in the from of the first law of
thermodynamics, $dE=T_hdS_h+WdV$, on the apparent horizon and vice
versa, one should consider the entropy expression of the black
hole in each gravity theory. The only change one should done is
replacing the black hole horizon radius $r_{+}$ by the apparent
horizon radius ${\tilde r}_A$. However, the entropy expression
associated with the black hole horizon get modified from the
inclusion of quantum effects. Several type of quantum corrections
to the area law have been introduced in the literatures, among
them are logarithmic and power-law corrections. Logarithmic
corrections, arises from the loop quantum gravity due to thermal
equilibrium fluctuations and quantum fluctuations
\cite{Log,Rovelli,Zhang}. The logarithmic term also appears in a
model of entropic cosmology which unifies the inflation and late
time acceleration \cite{YFCai}. Another form of correction to area
law, namely the power-law correction, appears in dealing with the
entanglement of quantum fields inside and outside the horizon
\cite{sau1,sau2,Sau,pavon1}. Another modification for the area law
of entropy comes from the Gibbs arguments who pointed out that in
systems with divergency in the partition function, like
gravitational system, the Boltzmann-Gibbs (BG) theory cannot be
applied. As a result thermodynamical entropy of such nonstandard
systems is not described by an additive entropy but must be
generalized to the non-additive entropy \cite{Tsa0}. Based on
this, and using the statistical arguments, Tsallis and Cirto
argued that the microscopic mathematical expression of the
thermodynamical entropy of a black hole does not obey the area law
and can be modified as \cite{Tsa},
\begin{eqnarray}\label{S}
S_{h}= \gamma A^{\beta},
\end{eqnarray}
where $A$ is the black hole horizon area, $\gamma$ is an unknown
constant and $\beta$ known as Tsallis parameter or nonextensive
parameter,  which is a real parameter which quantifies the degree
of nonextensivity \cite{Tsa}. It is obvious that the area law of
entropy is restored for $\beta=1$ and $\gamma=1/(4L_p^2)$. Through
this paper we set $k_{B}=1=c=\hbar $ for simplicity. In fact, at
this limit, the power-law distribution of probability becomes
useless, and the system is describable by the ordinary
distribution of probability \cite{Tsa,Barb}.

It is worth mentioning that in deriving Friedmann equations from
the first law of thermodynamics, the entropy expression associated
with the horizon plays a crucial role \cite{CaiLM}. Thus, it is
interesting to see how the Friedmann equation get modified if the
entropy-area relation gets corrections by some reasons. Starting
from the first law of thermodynamics at apparent horizon of a FRW
universe, and assuming that the associated entropy with apparent
horizon has a logarithmic quantum corrected relation, the modified
Friedmann equations were derived in \cite{SheyLog}. Also, taking
the associated entropy with apparent horizon as the
power-law-corrected relation, one is able to obtain the corrected
Friedmann equation by using the first law of thermodynamics at the
apparent horizon \cite{SheyPL}. Besides, if thermodynamical
interpretation of gravity near apparent horizon is generic
feature, one should also ont only check the first law of
thermodynamics but also the generalized second law of
thermodynamics. The latter is a universal principle governing the
evolution of the total entropy of the Universe. In the context of
the accelerating Universe, the generalized second law of
thermodynamics has been explored in
\cite{wang1,wang2,Shey3,akbar}. {It should be noted that dark
energy and a modified Friedmann equations in the context of
Tsallis entropy and from different perspective, were first
investigated in \cite{Barb,Nunes}. It was argued that Tsallis
entropy parameter change the strength of the gravitational
constant and consequently the energy density of the dark
components of the universe, requiring more (less) dark energy to
provide the observed late time universe acceleration \cite{Nunes}.
They also explored some phenomenological aspects as well as some
observational constraints from a modified Friedmann equations
induced by Tsallis entropy \cite{Barb,Nunes}. In the context of
the nonextensive Kaniadakis statistics \cite{Kan}, the Jeans
length was investigated and the results were compared with the
Jeans length obtained in the non-extensive Tsallis statistics
\cite{Abero}. Recently, modified cosmology through nonextensive
Tsallis entropy have been investigated in \cite{Em}. It was shown
that the universe exhibits the usual thermal history, with the
sequence of matter and dark energy eras, and depending on the
value of nonextensive parameter $\beta$ the equation of state of
dark energy can even cross the phantom-line \cite{Em}. In this
work, we shall derive the modified Friedmann equation in a
universe which its entropy is given by the nonextensive Tsallis
entropy. Then, we investigate the cosmological implications of the
obtained modified Friedmann equations in the matter and radiation
dominated era. In our study, we do not need to invoke the dark
energy component and the early deceleration  as well as the late
time acceleration of the universe expansion can be achieved in the
presence of radiation and pressureless matter. Besides, we shall
show that our model can solve the age problem of the universe. Our
approach and the obtained Friedmann equations, from the first law
of thermodynamics completely differ from those investigated in
\cite{Barb,Nunes,Em}.}

This paper is organized as follows. In the next section, we derive
the modified Friedmann equations by applying the first law of
thermodynamics, $dE=T_hdS_h+WdV$, at apparent horizon of a FRW
universe and assuming the entropy associated with apparent horizon
is in the form of Tsallis entropy (\ref{S}). In section \ref{GSL},
we examine the generalized second law of thermodynamics for the
total entropy including the corrected entropy-area relation
together with the matter field entropy inside the apparent
horizon. In section \ref{Eme}, we adopt the modified version of
Padmanabhan's proposal \cite{PadEm} for deriving the Friedmann
equation corresponding to Tsallis entropy (\ref{S}). By assuming
the difference between the number of degrees of freedom in the
bulk and on the boundary is proportional to the volume change of
the spacetime, we find the modified Friedmann equation which
coincides with the one obtained from the first law of
thermodynamics. {In section \ref{cosm}, we investigate the
cosmological consequences of the modified Friedmann equations
derived from the non-extensive Tsallis entropy.} The last section
is devoted to conclusion and discussion.
\section{Modified Friedman Equation from the First law of thermodynamics\label{FIRST}}
We assume the background spacetime is spatially homogeneous and
isotropic which is described by the line element
\begin{equation}
ds^2={h}_{\mu \nu}dx^{\mu}
dx^{\nu}+\tilde{r}^2(d\theta^2+\sin^2\theta d\phi^2),
\end{equation}
where $\tilde{r}=a(t)r$, $x^0=t, x^1=r$, and $h_{\mu \nu}$=diag
$(-1, a^2/(1-kr^2))$ represents the two dimensional metric. The
open, flat, and closed universes corresponds to $k = 0, 1, -1$,
respectively. We also assume the physical boundary of the
Universe, which is consistent with laws of thermodynamics, is the
apparent horizon with radius
\begin{equation}
\label{radius}
 \tilde{r}_A=\frac{1}{\sqrt{H^2+k/a^2}}.
\end{equation}
The associated temperature with the apparent horizon can be
defined as \cite{Cai2}
\begin{equation}\label{T}
T_h=\frac{\kappa}{2\pi}=-\frac{1}{2 \pi \tilde
r_A}\left(1-\frac{\dot {\tilde r}_A}{2H\tilde r_A}\right).
\end{equation}
where
 $\kappa$ is the surface gravity. For $\dot {\tilde r}_A\leq 2H\tilde r_A$, the temperature becomes $T\leq
0$. To avoid the negative temperature one may define
$T=|\kappa|/2\pi$.  Also, within an infinitesimal internal of time
$dt$  one may assume $\dot {\tilde r}_A\ll 2H\tilde r_A$, which
physically means that the apparent horizon radius is kept fixed.
Thus there is no volume change in it and one may define $T=1/(2\pi
\tilde r_A )$ \cite{CaiKim}. The profound connection between
temperature on the apparent horizon and the Hawking radiation has
been considered in \cite{cao}, which further confirms the
existence of the temperature associated with the apparent horizon.

We assume the matter and energy content of the Universe is in the
form of perfect fluid with stress-energy tensor
\begin{equation}\label{T1}
T_{\mu\nu}=(\rho+p)u_{\mu}u_{\nu}+pg_{\mu\nu},
\end{equation}
where $\rho$ and $p$ are the energy density and pressure,
respectively. The conservation of the energy-stress tensor in the
FRW background, $\nabla_{\mu}T^{\mu\nu}=0$, leads to the
continuity equation as
\begin{equation}\label{Cont}
\dot{\rho}+3H(\rho+p)=0,
\end{equation}
where $H=\dot{a}/a$ is the Hubble parameter. Following
\cite{Hay2}, we define the work density as
\begin{equation}\label{Work}
W=-\frac{1}{2} T^{\mu\nu}h_{\mu\nu}.
\end{equation}
A simple calculation gives
\begin{equation}\label{Work2}
W=\frac{1}{2}(\rho-p).
\end{equation}
The work density term is indeed the work done by the volume change
of the Universe, which is due to the change in the apparent
horizon radius. We assume the first law of thermodynamics on the
apparent horizon is satisfied and has the form
\begin{equation}\label{FL}
dE = T_h dS_h + WdV.
\end{equation}
It is clear that this equation is similar to the standard first
law of thermodynamics, unless the work term $-pdV$ is replaced by
$WdV$. For a pure de Sitter space where $\rho=-p$, the work term
reduces to the standard $-pdV$ and one arrives at the standard
first law of thermodynamics.

We suppose the total energy content of the universe inside a
$3$-sphere of radius $\tilde{r}_{A}$, is $E=\rho V$ where
$V=\frac{4\pi}{3}\tilde{r}_{A}^{3}$ is the volume enveloped by
3-dimensional sphere with the area of apparent horizon
$A=4\pi\tilde{r}_{A}^{2}$. Taking differential form of the total
matter and energy inside the apparent horizon, we find
\begin{equation}
\label{dE1}
 dE=4\pi\tilde
 {r}_{A}^{2}\rho d\tilde {r}_{A}+\frac{4\pi}{3}\tilde{r}_{A}^{3}\dot{\rho} dt.
\end{equation}
Using the continuity equation (\ref{Cont}), we obtain
\begin{equation}
\label{dE2}
 dE=4\pi\tilde
 {r}_{A}^{2}\rho d\tilde {r}_{A}-4\pi H \tilde{r}_{A}^{3}(\rho+p) dt.
\end{equation}
We further assume the entropy associated with the apparent horizon
is in the form of Tsallis entropy (\ref{S}), where now $A$ is the
apparent horizon area of the Universe. Differentiating the
modified entropy-area relation (\ref{S}), we get
\begin{equation} \label{dS}
dS_h= \gamma \beta A^{\beta-1}dA= 8\pi \gamma \beta (4 \pi
{r}_{A}^2 )^{\beta-1} \tilde {r}_{A} d\tilde {r}_{A}.
\end{equation}
Substituting Eqs. (\ref{Work2}), (\ref{dE2}) and (\ref{dS}) in the
first law (\ref{FL}) and using relation (\ref{T1}) we get the
differential form of the Friedmann equation as
\begin{equation} \label{Fried1}
\frac{\gamma \beta}{\pi \tilde {r}_{A}^3} \left(4\pi \tilde
{r}_{A}^2\right)^{\beta-1} d\tilde {r}_{A}= H(\rho+p) dt.
\end{equation}
Using the continuity equation (\ref{Cont}), we can rewrite it as
\begin{equation} \label{Fried2}
-\frac{2}{\tilde {r}_{A}^3} \left(4\pi \tilde
{r}_{A}^2\right)^{\beta-1} d\tilde {r}_{A} = \frac{2\pi }{3\gamma
\beta}d\rho.
\end{equation}
Next, we integrate Eq. (\ref{Fried2}),
\begin{equation} \label{Fried3}
-2\left(4\pi \right)^{\beta-1}\int{\tilde {r}_{A}^{2\beta-5}
d\tilde {r}_{A}} = \frac{2\pi }{3\gamma \beta} \rho,
\end{equation}
which can be written as
\begin{equation} \label{Frie3}
\frac{1}{\tilde {r}_{A}^{4-2\beta}}= \frac{2\pi (2-\beta)
}{3\gamma \beta} \left(4\pi \right)^{1-\beta}  \rho,
\end{equation}
where we have set the integration constant equal to zero.
Substituting $\tilde {r}_{A}$ from Eq.(\ref{radius}) we obtain
\begin{equation} \label{Fried4}
\left(H^2+\frac{k}{a^2}\right)^{2-\beta} = \frac{8\pi L_p^2} {3}
\rho,\end{equation} provided we define
\begin{equation}\label{Lp}
\gamma\equiv\frac{2-\beta }{4\beta L_p^2 } \left(4\pi
\right)^{1-\beta}.
\end{equation}
Equation (\ref{Fried4}) is nothing but the modified Friedmann
equation obtained through the non-additive Tsallis entropy. In
this way we have not only derived the modified Friedmann equation
by starting from the first law of thermodynamics at apparent
horizon of a FRW universe, and assuming that the associated
entropy with apparent horizon has a corrected relation (\ref{S}),
but also find a general definition for the unknown constant
$\gamma$ in the entropy expression (\ref{S}). In the limiting case
where $\beta=1$, we recover the standard Friedmann equation in
Einstein gravity. Besides, from Eq. (\ref{Lp}), we have always
$\beta<2$ which puts an upper bound on the non-additive parameter
of the Tsallis entropy.
\section{Generalized Second law of thermodynamics\label{GSL}}
Now we want to examine the validity of the generalized second law
of thermodynamics in a region enclosed by the apparent horizon.
Combining Eq. (\ref{Fried2}) with Eq. (\ref{Cont}) and using
relation (\ref{Lp}), we arrive at
\begin{equation} \label{GSL1}
\frac{2}{\tilde {r}_{A}^3} (2-\beta) \dot{\tilde {r}}_{A} \tilde
{r}_{A}^{2\beta-2}=8\pi L_p^2 H (\rho+p).
\end{equation}
Solving for $\dot{\tilde {r}}_{A}$ we find
\begin{equation} \label{dotr1}
\dot{\tilde {r}}_{A}=\frac{4 \pi L_p^2 H}{2-\beta} \tilde
{r}_{A}^{5-2\beta}(\rho+p).
\end{equation}
Since $\beta<2$, thus the sign of $\dot{\tilde {r}}_{A}$ depends
on the sign of $\rho+p$. For $\rho+p>0$, which physically means
the dominant energy condition holds, we have
$\dot{\tilde{r}}_{A}>0$. Let us now turn to find out $T_{h}
\dot{S_{h}}$:
\begin{equation}\label{TSh1}
T_{h} \dot{S_{h}} =\frac{1}{2\pi \tilde r_A}\left(1-\frac{\dot
{\tilde r}_A}{2H\tilde r_A}\right)\frac{d}{dt}\left[\gamma (4 \pi
\tilde {r}_{A}^2)^\beta \right].
\end{equation}
After some simplifications and using Eq. (\ref{dotr1}) we obtain
\begin{equation}\label{TSh2}
T_{h} \dot{S_{h}} =4\pi H {\tilde{r}_{A}^3}
(\rho+p)\left(1-\frac{\dot {\tilde r}_A}{2H\tilde r_A}\right).
\end{equation}
At present, our Universe is undergoing an acceleration phase and
thus its equation of state parameter can cross the phantom line
($w_D=p/\rho<-1$). This implies that in an accelerating universe
the dominant energy condition may violate, $\rho+p<0$, implying
that the second law of thermodynamics, $\dot{S_{h}}\geq0$, does
not hold. Therefore, we should examine the validity of the
generalized second law of thermodynamics, namely
$\dot{S_{h}}+\dot{S_{m}}\geq0$.

From the Gibbs equation we have \cite{Pavon2}
\begin{equation}\label{Gib2}
T_m dS_{m}=d(\rho V)+pdV=V d\rho+(\rho+p)dV,
\end{equation}
where $T_{m}$ and $S_{m}$ are, respectively, the temperature and
the entropy of the matter fields inside the apparent horizon. We
assume the local equilibrium hypothesis holds. Therefore, the
thermal system bounded by the apparent horizon remains in
equilibrium so that the temperature of the system must be uniform
and the same as the temperature of its boundary, $T_m=T_h$
\cite{Pavon2}. Note that if the temperature of the fluid differs
much from that of the horizon, there will be spontaneous heat flow
between the horizon and the bulk fluid and the local equilibrium
hypothesis will no longer hold. Thus, from the Gibbs equation
(\ref{Gib2}) we can obtain
\begin{equation}\label{TSm2}
T_{h} \dot{S_{m}} =4\pi {\tilde{r}_{A}^2}\dot {\tilde
r}_A(\rho+p)-4\pi {\tilde{r}_{A}^3}H(\rho+p).
\end{equation}
In order to examine the generalized second law of thermodynamics,
we should study the evolution of the total entropy $S_h + S_m$.
Adding equations (\ref{TSh2}) and (\ref{TSm2}),  we get
\begin{equation}\label{GSL2}
T_{h}( \dot{S_{h}}+\dot{S_{m}})=2\pi{\tilde r_A}^{2}(\rho+p)\dot
{\tilde r}_A=\frac{A}{2}(\rho+p) \dot {\tilde r}_A.
\end{equation}
where $A$ is the apparent horizon area. Inserting $\dot {\tilde
r}_A$ from Eq. (\ref{dotr1}) into (\ref{GSL2}) we find
\begin{equation}\label{GSL3}
T_{h}( \dot{S_{h}}+\dot{S_{m}})=\frac{8\pi^2}{2-\beta} L_p^2 H
{\tilde r_A}^{7-2\beta}(\rho+p)^2.
\end{equation}
Since $\beta<2$, the right hand side of the above equation is
always a non-negative function during the universe history, which
means that $ \dot{S_{h}}+\dot{S_{m}}\geq0$. This implies that for
a universe with Tsallis entropy the total entropy namely the
entropy of the boundary together with the matter entropy inside
the bulk is a non decreasing function of time and hence the
generalized second law of thermodynamics is fulfilled.
\section{Modified Friedmann equation from emergence of cosmic space\label{Eme}}
According to the Padmanabhan's proposal \cite{PadEm}, the basic
equation governing the evolution of the Universe can be derived by
relating the emergence of space to the difference between the
number of degrees of freedom in the holographic surface and the
one in the emerged bulk. In this viewpoint, the spatial expansion
of our Universe can be regarded as the consequence of emergence of
space and  the cosmic space is emergent, following the progressing
in the cosmic time. Thus, he argued that in an infinitesimal
interval $dt$ of cosmic time, the increase $dV$ of the cosmic
volume, is given by  \cite{PadEm}
\begin{equation}
\frac{dV}{dt}=L_{p}^{2}(N_{\mathrm{sur}}-N_{\mathrm{bulk}}),
\label{dV}
\end{equation}
where $N_{\mathrm{sur}}$ is the number of degrees of freedom on
the boundary and $N_{\mathrm{bulk}}$ is the number of degrees of
freedom in the bulk. Using this new idea, Padmanabhan derived the
Friedmann equation of a flat FRW Universe \cite{PadEm}. It was
argued that this proposal failed to derive the Firedmann equations
of a nonflat FRW universe in other gravity theories such as
Gauss-Bonnet and Lovelock gravity \cite{CaiEm}. The modification
of Padmanabhan's proposal was done by the present author
\cite{Sheyem}, who argued that in a nonflat Universe the proposal
should be generalized as
\begin{equation}
\frac{dV}{dt}=L_{p}^{2}\frac{\tilde{r}_A}{H^{-1}}
\left(N_{\mathrm{sur}}-N_{\mathrm{bulk}}\right). \label{dV1}
\end{equation}
This implies that the volume increase, in a nonflat universe, is
still proportional to the difference between the number of degrees
of freedom on the apparent horizon and in the bulk, but the
function of proportionality is not just a constant, and is equal
to the ratio of the apparent horizon and Hubble radius. Clearly,
for spatially flat universe, $\tilde{r}_A =H^{-1}$, and one
recovers the proposal (\ref{dV}).

Now, we would like to see whether one can derive the modified
Friedmann equation from the relation (\ref{dV1}), when the entropy
associated with the apparent horizon get modified to Tsallis
entropy. First of all, we define the effective area of the
holographic surface corresponding to the entropy (\ref{S}) as
\begin{eqnarray}
\widetilde{A} = A^{\beta}= \left(4 \pi
{\tilde{r}_A^2}\right)^{\beta}.
\end{eqnarray}
Next, we calculate the increasing in the effective volume as
\begin{eqnarray}\label{dVt1}
\frac{d\widetilde{V}}{dt}&=&\frac{\tilde{r}_A}{2}\frac{d\widetilde{A}}{dt}=\beta
(4 \pi \tilde{r}_A^{2})^{\beta} \dot{\tilde{r}}_A\nonumber\\
&=&\beta (4 \pi)^{\beta} \frac{\tilde{r}_A^5}{(2\beta-4)} \
\frac{d}{dt}\left(\tilde{r}_A ^{2\beta-4}\right) . \label{dVt2}
\end{eqnarray}
Inspired by (\ref{dVt2}), we propose that the number of degrees of
freedom on the apparent horizon with Tsallis entropy, is given by
\begin{equation}
N_{\mathrm{sur}}=\frac{4\gamma \beta}{2-\beta} (4 \pi
\tilde{r}_A^2)^{\beta}. \label{Nsur2}
\end{equation}
We also assume the temperature associated with the apparent
horizon is the Hawking temperature, which is given by
\cite{CaiKim}
\begin{equation}\label{T2}
T=\frac{1}{2\pi \tilde{r}_A},
 \end{equation}
and the energy contained inside the sphere with volume $V=4
\pi\tilde{r}^3_A/3$ is the Komar energy
\begin{equation}
E_{\mathrm{Komar}}=|(\rho +3p)|V.  \label{Komar}
\end{equation}
According to the equipartition law of energy, the bulk degrees of
freedom obey
\begin{equation}
N_{\mathrm{bulk}}=\frac{2|E_{\mathrm{Komar}}|}{T}.  \label{Nbulk}
\end{equation}
In order to have $N_{\rm bulk}>0$, we take $\rho+3p<0$
\cite{PadEm}. Thus the number of degrees of freedom in the bulk is
given by
\begin{equation}
N_{\rm bulk}=-\frac{16 \pi^2}{3}  \tilde{r}_A^{4} (\rho+3p),
\label{Nbulk}
\end{equation}
We also replace $L_p^2$ with $1/(4\gamma)$ and $V$ with
$\widetilde{V}$ in  proposal (\ref{dV1}) and write it down as
\begin{equation}
4\gamma \frac{d\widetilde{V}}{dt}=\frac{\tilde{r}_A}{H^{-1}}
(N_{\mathrm{sur}}-N_{\mathrm{bulk}}). \label{dV2}
\end{equation}
Substituting relations (\ref{dVt1}), (\ref{Nsur2}) and
(\ref{Nbulk}) in Eq. (\ref{dV2}), after simplifying, we arrive at
\begin{eqnarray}
(2-\beta)\tilde{r}_A^{2\beta-5} \
\frac{\dot{\tilde{r}}_A}{H}-\tilde{r}_A^{2\beta-4} &=&\frac{4 \pi
L_{p}^{2}}{3}(\rho+3p), \label{Frgb1}
\end{eqnarray}
where we have also used  definition (\ref{Lp}).  Multiplying the
both hand side of Eq. (\ref{Frgb1}) by factor $2\dot{a}a$, after
using the continuity equation (\ref{Cont}), we reach
\begin{equation}
\frac{d}{dt}\left( a^2 \tilde{r}_A^{2\beta-4}\right)=\frac{8 \pi
L_{p}^{2}}{3} \frac{d}{dt}(\rho a^2). \label{Frgb2}
\end{equation}
Integrating, yields
\begin{equation}
\left(H^2+\frac{k}{a^2}\right)^{2-\beta}=\frac{8 \pi L_{p}^{2}}{3}
\rho, \label{Frgb3}
\end{equation}
where we have set the integration constant equal to zero. This is
nothing but the modified Friedmann equation motivated from Tsallis
entropy-area relation (\ref{S}). We see that our result from the
emergence approach coincides with the modified Friedmann equation
derived from the first law of thermodynamics in section
\ref{FIRST}. Our study indicates that the approach presented here
is enough powerful and further supports the viability of the
Padmanabhan's perspective of emergence gravity and its
modification given by Eq. (\ref{dV2}).

\section{Tsallis Cosmology}\label{cosm}
In this section, we would like to investigate cosmological
consequences of the modified Friedmann equation derived in Eqs.
(\ref{Fried4}) and (\ref{Frgb3}). Let us begin by deriving the
second modified Friedmann equation in Tsallis cosmology. Taking
the time derivative of the first Friedmann equation (\ref{Frgb3}),
we get
\begin{equation}
2H(2-\beta) \left(\dot{H}-
\frac{k}{a^2}\right)\left(H^2+\frac{k}{a^2}\right)^{1-\beta}=\frac{8
\pi L_{p}^{2}}{3} \dot{\rho}. \label{2Fri1}
\end{equation}
Using the continuity equation (\ref{Cont}), we arrive at
\begin{eqnarray}
&&(2-\beta) \left(\dot{H}-
\frac{k}{a^2}\right)\left(H^2+\frac{k}{a^2}\right)^{1-\beta}=-4\pi
L_{p}^{2} (\rho+p) \nonumber \\&&
-\frac{3}{2}\left(H^2+\frac{k}{a^2}\right)^{2-\beta}-4\pi
L_{p}^{2} p,\label{2Fri2}
\end{eqnarray}
where we have also used Eq. (\ref{Frgb3}) in the last step. Now
using the  fact that $\dot{H}=\ddot{a}/a-H^2$, after some
calculations, we can rewrite the above equation as
\begin{eqnarray}
(4-2\beta)
\frac{\ddot{a}}{a}+(2\beta-1)\left(H^2+\frac{k}{a^2}\right)^{2-\beta}=-8\pi
L_{p}^{2} p.\label{2Fri3}
\end{eqnarray}
This is indeed the second modified Friedmann equation governing
the evolution of the Universe in Tsallis cosmology which is based
on the nonextensive Tsallis entropy. When $\beta=1$, the above
equation reduces to the second Friedmann equation in standard
cosmology,
\begin{eqnarray}
2 \frac{\ddot{a}}{a}+\left(H^2+\frac{k}{a^2}\right)=-8\pi
L_{p}^{2} p.\label{2Fri4}
\end{eqnarray}
We can also obtain the equation for the second time derivative of
the scale factor. To this aim, we combine the first and second
modified Friedmann equations (\ref{Frgb3}) and (\ref{2Fri3}).  We
find

\begin{eqnarray}
\frac{\ddot{a}}{a}=-\frac{4\pi L_{p}^{2}\rho}{3(2-\beta)}
\left[(2\beta-1) \rho +3p\right]. \label{2Fri5}
\end{eqnarray}
Since our Universe is currently undergoing an acceleration phase
($\ddot{a}>0$), thus from the above equation we should have
\begin{eqnarray}
 (2\beta-1) \rho +3p <0  \  \     \longrightarrow   \  \  \omega< \frac{1-2\beta}{3},          \label{w1}
\end{eqnarray}
where $\omega=p/\rho$ is the equation of state parameter. Again,
for $\beta=1$ the above condition for the accelerated universe
leads to the well-known inequality $\omega<-1/3$ for the equation
of state in standard cosmology. A close look on the inequality
(\ref{w1}) show that for $\beta\geq1/2$, we should always have
$\omega< 0$. However,  for $\beta<1/2$, we can have $\omega\geq0$
in an accelerated universe. For example, for $\beta=1/3$ the above
inequality implies $\omega<1/9$. This is an interesting result,
which indicates that in Tsallis cosmology, the late time
accelerated universe can be achieved even in the presence of the
ordinary matter with positive equation of state parameter. Thus,
if we assume our Universe is now dominated with pressureless
matter with ($\omega=0$), then it can undergo an accelerated
expansion provided we choose the nonextensive parameter $\beta$
less than $1/2$, without needing to an additional component of
energy such as dark energy.

Next, we are going to find the scalae factor as a function of time
in Tsallis cosmology. For simplicity we only consider the flat
universe with $k=0$, although the calculations can be easily
extended to the nonflat universe ($k\neq0$). We shall consider the
radiation and matter dominated era, separately.

\subsection{Matter-dominated era}
It is well-known that the pressure caused by random motions of
galaxies and clusters of galaxies is negligible. This is due to
the fact that the random velocities of the galaxies are of the
order of $10^8$ $\rm cm \rm s^{-1}$ or less \cite{GRbook}.
Besides, the average energy density of the universe is of the
order of $10^{-30} \rm g \rm cm^{-3}$. Thus, we can write
\begin{eqnarray}
 p\sim \rho <v^2> \sim 10^{-5} \rho c^2 \ll \rho c^2.
\end{eqnarray}
This implies that the pressure is negligible compared with $\rho
c^2$. Thus if the energy density of the universe is dominated by
the nonrelativistic matter with negligible pressure, we can write
$p=0$. In this case, the continuity equation (\ref{Cont}) can be
written $\dot{\rho}(t)+3H\rho(t)=0$, which can be easily
integrated to yield
\begin{eqnarray} \label{rhom}
\frac{\rho(t)}{\rho(t_0)}=\left(\frac{a(t)}{a(t_0)}\right)^{-3},
\label{rho}
\end{eqnarray}
where $t_0$ is the present time of the universe. Therefore, $\rho
a^3=\rm constant.$, which is exactly the same as in standard
cosmology in the matter dominated era. Substituting $\rho$ from
(\ref{rhom}) in the first Friedmann equation (\ref{Frgb3}), we
arrive at
\begin{eqnarray}
\left(\frac{\dot{a}}{a}\right)^{4-2\beta}= \frac{8 \pi L_{p}^{2}
\rho_0 a_0^3}{3 a^3},
\end{eqnarray}
which can be rewritten as
\begin{eqnarray} \label{adot}
\left(\frac{d{a}}{dt}\right)^{4-2\beta}= C_1 a^{1-2\beta},
\end{eqnarray}
where we have defined
\begin{eqnarray}
C_1\equiv\frac{8 \pi L_{p}^{2} \rho_0 a_0^3}{3}.
\end{eqnarray}
Eq. (\ref{adot}) has the following solution,
\begin{eqnarray}
a(t)= (C_2 t)^{(4-2\beta)/3},
\end{eqnarray}
where
\begin{eqnarray}
C_2\equiv \frac{3}{2} \frac{ C_1 ^{1/(4-2\beta)} }{(2-\beta)}>0 .
\end{eqnarray}
In the limit of standard cosmology where $\beta=1$, the above
solution restores
\begin{eqnarray}
a(t)= \left[\frac{3}{2}\sqrt{C_1}\right]^{2/3} t^{2/3},
\end{eqnarray}
Let us now calculate the second time derivative of the scale
factor.  We find
\begin{eqnarray}
\ddot{a}(t)=  \frac{C_2^{(4-2\beta)/3}}{9}(4-2\beta)(1-2\beta) \
t^{-(2+2\beta)/3},
\end{eqnarray}
Again, we see that in the matter dominated universe, the
accelerated expansion ($\ddot{a}(t)>0$) can be achieved provided
we take $\beta<1/2$, which is consistent with our previous
discussion. Thus, in the framework of Tsallis cosmology, without
invoking any kind of dark energy and only with pressureless
matter, the late time acceleration of the universe expansion can
be understood.

We can also calculate the evolution of the energy density, the
Hubble and the deceleration parameters as
\begin{eqnarray}
\rho(t)& \propto & \frac{1}{t^{4-2\beta}},\\
H(t)&=& \frac{\dot{a}}{a}=\frac{4-2\beta}{3 t},\\
q(t)&=& -\frac{a\ddot{a}}{\dot{a}^2}=\frac{2\beta-1}{4-2\beta}.
\end{eqnarray}
Again, all above parameters are reduced to the one in standard
cosmology for $\beta=1$. From the deceleration parameter, we see
that for $\beta<1/2$, we have $q<0$ and $\ddot {a}>0$, a result
which confirm the accelerated universe. Now, we look at the Hubble
parameter which can be employed for estimating the age of the
universe at the present time ($t=t_0$). We have
\begin{equation}\label{age}
t_0=\frac{4-2\beta}{3H_0},
\end{equation}
where $H_0=H(t_0)$ is the Hubble constant. In an accelerated
universe we have $\beta<1/2$, which implies $4-2\beta>3$, and thus
\begin{equation}\label{age1}
t_0>\frac{1}{H_0}=\frac{3}{2}\left(\frac{2}{3H_0}\right),
\end{equation}
Note that $2/(3H_0)$ is the age of the universe in standard
cosmology. Let us see how the above relation can alleviate the
problem of age in standard cosmology. The Hubble constant is
usually written as
\begin{equation}
{H_0}=100  h \  \rm km \ \rm sec^{-1} \rm Mpc=2.1332  \ h \times
10^{-42} \rm GeV.
\end{equation}
where $h$ describes the uncertainty on the value $H_0$. The
observations of the Hubble Key Project constrain this value to be
$h=0.72\pm 0.08$ \cite{age1}. Thus the Hubble time is
$t_H=1/H_0=9.78 \times 10^{9} h^{-1} \rm years$. Using this value
for $h$, the age of the Universe in standard cosmology is in the
range $8.2$ Gyr $< t_0 < 10.2 $ Gyr  \cite{DEbook}. Carretta et
al. \cite{age2} estimated the age of globular clusters in the
Milky Way to be $12.9 \pm 2.9 $ Gyr, whereas Jimenez et al.
\cite{age3} obtained the value $13.5\pm2 $ Gyr. Hansen et al.
\cite{age4} constrained the age of the globular cluster $M4$ to be
$12.7 \pm 0.7$ Gyr by using the method of the white dwarf cooling
sequence. In most cases the ages of globular clusters are larger
than $11$ Gyr which indicates that the cosmic age estimated in
standard cosmology is inconsistent with the ages of the oldest
globular clusters. It was argued that, in standard model of
cosmology, this problem cannot be circumvented unless the
cosmological constant (dark energy) is taken into account
\cite{DEbook}.

However, in the framework of Tsallis cosmology,  the problem of
age can be circumvented automatically for an accelerated universe.
More precisely, from relation (\ref{age1}) the age of the universe
in Tsallis cosmology is larger than $3/2$ age of the universe in
standard cosmology, namely
\begin{equation}\label{age2}
t_0|_{T}>\frac{3}{2} t_0|_{S},
\end{equation}
provided we choose $\beta<1/2$. Here subscript ``T" and ``S" stand
for Tsallis and Standard cosmology, respectively. As an example,
taking $\beta=2/5$, from relation (\ref{age}) we find
$t_0|_{T}=1.6 \ t_0|_{S}$. Consequently, the age of the
accelerated universe in Tsallis cosmology is in the range $13.12$
Gyr $< t_0|_{T} < 16.32 $ Gyr, which is larger than the age of the
oldest globular clusters. Therefore, the cosmic age problem can be
properly alleviated in an accelerated universe in the framework of
Tsallis cosmology without invoking additional component of energy.

\subsection{Radiation-dominated era}
It is well-known that, in an expanding universe, the proper
momenta of freely moving particles decreases in term of scale
factor as $1/a(t)$. Thus, small random velocities of particles
seen today should have been large in the past when the scale
factor was much smaller than its present value \cite{GRbook}. As a
result, the pressureless approximation is break down in the early
universe. In this subsection, we are going to find the solution
for the modified Friedmann equation in the framework of Tsallis
cosmology by assuming that the universe is filled with a highly
relativistic gas (radiation) with equation of state $p=\rho/3$. In
this case from the continuity equation,
$\dot{\rho}(t)+4H\rho(t)=0$, we can get
\begin{eqnarray} \label{rhor}
\frac{\rho(t)}{\rho(t_1)}=\left(\frac{a(t)}{a(t_1)}\right)^{-4},
\label{rho}
\end{eqnarray}
where $t_1$ is an arbitrary reference time in the radiation
dominated era. Thus, in this case $\rho a^4=\rm constant.$, which
is similar to the result obtained in the radiation dominated era
in standard cosmology. Substituting $\rho$ from (\ref{rhor}) in
the first Friedmann equation (\ref{Frgb3}), we get
\begin{eqnarray}
\left(\frac{\dot{a}}{a}\right)^{4-2\beta}= \frac{8 \pi L_{p}^{2}
\rho_1 a_1^4}{3 a^4},
\end{eqnarray}
where $\rho_1=\rho(t_1)$, $a_1=a(t_1)$ and we have again
considered the flat universe ($k=0$), although the $k=\pm1$ cases
can also be easily solved. The above equation can be rewritten as
\begin{eqnarray} \label{adotr}
\left(\frac{d{a}}{dt}\right)^{4-2\beta}= B_1 a^{-2\beta},
\end{eqnarray}
where
\begin{eqnarray}
B_1\equiv\frac{8 \pi L_{p}^{2} \rho_1 a_1^4}{3}.
\end{eqnarray}
Eq. (\ref{adotr}) admits a solution of the form,
\begin{eqnarray}
a(t)= B_1^{1/4} \left(\frac{2 t}{2-\beta}\right)^{1-\beta/2}.
\end{eqnarray}
When $\beta=1$, we arrive at
\begin{eqnarray}
a(t)= \sqrt{2}B_1^{1/4} t^{1/2},
\end{eqnarray}
which is the corresponding solution in standard cosmology. The
second time derivative of the scale factor is obtained as
\begin{eqnarray}
\ddot{a}(t)= -\frac{\beta}{2} B_1^{1/4}
\left(\frac{2}{2-\beta}\right)^{-\beta/2} t^{-(1+\beta/2)}<0.
\end{eqnarray}
This shows that in the radiation dominated era, the universe has
been in an decelerated phase ($\ddot{a}(t)<0$). Now, we calculate
the energy density, the Hubble and the deceleration parameters in
the radiation dominated era. We find
\begin{eqnarray}
\rho(t)& \propto & \frac{1}{t^{4-2\beta}},\\
H(t)&=& \frac{\dot{a}}{a}=\frac{2-\beta}{2 t},\\
q(t)&=& -\frac{a\ddot{a}}{\dot{a}^2}=\frac{\beta}{2-\beta}.
\end{eqnarray}
Therefore, for $\beta<2$, we have always $q>0$ ($\ddot{a}(t)<0$),
which confirms that we have a decelerated universe in the
radiation dominated era. This implies that in the early stage of
the universe where the relativistic particles have been dominated,
our Universe has been in a decelerated phase and at the late time
it undergoes an accelerated expansion. In summary, Tsallis
cosmology can explain the evolution of the universe from the early
deceleration to the late time acceleration, without invoking dark
energy, which is consistent with recent observations.

\subsection{Density perturbation}
In order to study  the density perturbation, we follow the method
which was first introduced by James Jeans \cite{GRbook}.  Assuming
a flat FRW universe in the matter-dominated era, the fundamental
differential equation governing the fractional density
perturbation $\delta=\rho_1/\rho$ in an expanding universe becomes
\cite{GRbook}

\begin{eqnarray}\label{delta1}
\ddot{\delta}+\frac{2 \dot{a}}{a} \dot{\delta}+\left(\frac{v_s^2
\kappa^2}{a^2}-4\pi G \rho\right) \delta=0,
\end{eqnarray}
where $v_s^2=p_1/\rho_1$ is the speed of sound, $\rho_1$ and $p_1$
stand for the perturbed part of the energy density and pressure,
respectively, and $\kappa$ is the co-moving wavevector. The
solution of the above equation depends on the sign of the function
\begin{eqnarray}
f(t)=\frac{v_s^2 \kappa^2}{a^2(t)}-4\pi G \rho (t).
\end{eqnarray}
For $f(t)>0$, the solution for $\delta(t)$ is periodic and the
perturbation propagate in the medium as sound waves. For $f(t)<0$,
however, lead to imaginary frequency, indicating a growth of
$\delta(t)$ with time. In what follows we are interested in the
second case where $f(t)<0$ and further assume $\frac{v_s^2
\kappa^2}{a^2}\ll 4\pi G \rho $. In this case Eq. (\ref{delta1})
can be written
\begin{eqnarray}\label{delta2}
\ddot{\delta}+\frac{2 \dot{a}}{a} \dot{\delta}-4\pi G \rho
\delta=0.
\end{eqnarray}
In the framework of standard model of cosmology where $a(t)\propto
t^{2/3}$, Eq. (\ref{delta2}) can be further rewritten as
\begin{eqnarray}\label{delta3}
\ddot{\delta}+\frac{4}{3t} \dot{\delta}-\frac{2}{3t^2} \delta=0.
\end{eqnarray}
which has a solution in the form $\delta(t)=\delta_0 t^{2/3}$, and
thus the perturbation grows with time. On the other hand, when the
Friedmann equations are governed by the Tsallis cosmology, the
scale factor in the matter dominated era is, $a(t)\propto
t^{(4-2\beta)/3}$. On substituting in Eq. (\ref{delta2}), we
arrive at
\begin{eqnarray}\label{delta3}
\ddot{\delta}+\frac{4}{3t} (2-\beta)\dot{\delta}-\frac{3}{2}
\left(\frac{4-2\beta}{3t}\right)^{4-2\beta} \delta=0.
\end{eqnarray}
The solution of the above equation is given by
\begin{eqnarray}\label{delta4}
&&\delta(t)=\delta_0 \times  \nonumber\\
&& BesselJ \left(\frac{5-4 \beta}{6\beta-6},
\frac{2(\beta-2)^2}{9\beta-9}
\sqrt{-6\left(\frac{4-2\beta}{3}\right)^{-2\beta}}
t^{\beta-1}\right),\nonumber \\
\end{eqnarray}
To have an insight on the form of $\delta(t)$, let us consider a
special case where $\beta=5/4$ at the early stage of the universe.
In this case the expansion of solution (\ref{delta4}) is given by
\begin{eqnarray}\label{deltaexp}
&&\delta(t)=\delta_0 \left[ 1-\alpha_1 t^{1/2}+\alpha_2 t + ...
\right]
\end{eqnarray}
where $\alpha_i$ are constants. Thus, the growth of energy differs
in Tsallis cosmology compared to the standard cosmology. It is
important to note that this study is very brief and we leave the
details of investigations on the density perturbation in Tsallis
cosmology for future studies.
\section{conclusion and discussion\label{Con}}
It was already pointed out by Gibbs in 1902 that in systems whose
partition function diverges, like gravitational system, the
Boltzmann-Gibbs (BG) theory cannot be applied. As a result, the
thermodynamical entropy of such nonstandard systems is not
described by an additive entropy but with appropriately
generalized nonadditive entropies. Based on this, and using the
statistical arguments, Tsallis and Cirto \cite{Tsa} argued that
the microscopic mathematical expression of the thermodynamical
entropy of a black hole does not obey the area law and can be
modified as in Eq. (\ref{S}). Besides, it is well-known that the
entropy of the whole universe, considered as a system with
apparent horizon radius has a similar expression to the entropy
associated with black hole horizon. The only change is needed
replacing the black hole horizon  radius $r_{+}$ by the apparent
horizon radius $\tilde{r}_A$ of the Universe.

In this paper, we have assumed the entropy associated with the
apparent horizon of FRW universe is in the form of the Tsallis
entropy and examined its consistency with the laws of
thermodynamics. For this purpose, we first assumed the the first
law of thermodynamics, $dE=T_hdS_h+WdV$, holds on the apparent
horizon and the Tsallis entropy has the form (\ref{S}). Then, we
showed that the first law of thermodynamics on the apparent
horizon can be rewritten in the form of the modified Friedmann
equations of a FRW universe with any spatial curvature. We have
also examined the generalized second law of thermodynamics, by
studying the time evolution of the total entropy including the
entropy of the matter and energy inside the Universe together with
the Tsallis entropy associated with the apparent horizon. Assuming
the local equilibrium hypothesis, we confirmed that the
generalized second law of thermodynamics is fulfilled in a region
enclosed by the apparent horizon.

We have also considered the idea of emergence of cosmic space
proposed by Padmanabhan \cite{PadEm} and its modification given by
Eq. (\ref{dV2}) \cite{Sheyem}. Assuming the entropy associated
with the apparent horizon in the form of Tsallis entropy, we
determine the number of degrees of freedom on the boundary. Then,
by calculating the difference between the degrees of freedom on
the boundary, $N_{\rm sur}$, and the degrees of freedom in the
bulk, $N_{\rm bulk}$, we showed that one can extract the Friedmann
equations corresponding to the Tsallis entropy-area relation
(\ref{S}) which coincides with the one obtained from the first law
of thermodynamics. This study strongly supports the viability of
the novel idea proposed in \cite{PadEm,Sheyem}. Note that, we have
only modified the number of degrees of freedom on the boundary due
to the change in the entropy expression, while we assumed the bulk
degrees of freedom, $N_{\rm bulk}$, has the same expression as in
standard cosmology. The reason comes from the fact that we have
assumed $N_{\rm bulk}$ depends only on the matter degrees of
freedom, while $N_{\rm sur}$ crucially depends on the entropy
expression or/and underlying theory of gravity.

{We have also investigated some cosmological consequences of the
modified Friedmann equations. For this purpose, we first derived
the second modified Friedmann equation as well as the equation for
the second time derivative of the scale factor. Then, we
considered the matter-dominated era and the radiation-dominated
era, separately. We found that, in the matter dominated era which
is filled with pressureless matter, this model admits an
accelerated universe ($\ddot{a}(t)>0$) provided $\beta<1/2$.
Besides, the problem of age of universe can be circumvented
automatically for an accelerated universe. By calculating the
Hubble parameter in the matter dominated era, we estimated the age
of the universe larger than $3/2$ age of the universe in standard
cosmology. For example, taking $\beta=2/5$, we estimated the age
of the accelerated universe in Tsallis cosmology in the range
$13.12$ Gyr $< t_0|_{T} < 16.32 $ Gyr, which is larger than the
age of the oldest globular clusters. Therefore, the cosmic age
problem can be alleviated in an accelerated universe in the
framework of Tsallis cosmology without invoking additional
component of energy. In the radiation dominated era we have always
$q>0$ ($\ddot{a}(t)<0$), which confirms the deceleration phase for
the early stage of the universe. Therefore, Tsallis cosmology can
explain the evolution of the universe from the early deceleration
to the late time acceleration which is consistent with recent
observations. Finally, employing the Jeans's analysis, we explored
in brief, density perturbation in the matter dominated era in the
context of Tsallis cosmology. We showed that due to the
modification of the Friedmann equations and hence the scale
factor, the growth of energy is changed compared to the standard
cosmology.}

It is important to note that many issues remain to be studied.
First, the details of the effects of the modified Friedmann
equations on the growth of energy density in the early stage of
the universe as well as its influences on the structure formation
deserve further investigation. It is also interesting to consider
other forms of entropies e.g., the generalized Kaniadakis entropy
\cite{Kan} to investigate the modified Friedmann equations as well
as its cosmological consequences. These issues are under
investigations and the results will be appeared in the future
works.

\acknowledgments{I am grateful to the referee for very
constructive comments which helped me improve the paper
significantly. I also thank Shiraz University Research Council.
This work has been supported financially by Research Institute for
Astronomy and Astrophysics of Maragha (RIAAM), Iran.}


\end{document}